\documentclass[sigconf]{acmart}
\AtBeginDocument{%
  }

\definecolor{wsorange}{RGB}{245,166,115}

\newcommand{\parh}[1]{\noindent\textbf{#1}}

\usepackage{threeparttable}
\usepackage{xspace}
\usepackage{amsmath,amsfonts}
\usepackage{graphicx}
\usepackage{textcomp}
\usepackage{xcolor}
\usepackage{threeparttable}
\usepackage{colortbl}
\usepackage{graphicx}
\usepackage{listings}
\usepackage{color}
\usepackage{array}
\usepackage{float}
\usepackage{graphicx}
\usepackage{multirow}
\usepackage{colortbl}
\usepackage[ruled,linesnumbered,boxed]{algorithm2e}
\usepackage{algpseudocode}
\usepackage{pifont}
\usepackage{enumitem}
\usepackage{balance}
\usepackage{xurl}
\usepackage{hyperref}
\usepackage[T1]{fontenc}
\usepackage[utf8]{inputenc}
\usepackage{caption}
\usepackage{booktabs}
\usepackage{tabularx}
\usepackage{diagbox}
\usepackage{amsmath}
\usepackage{listings}
\usepackage{makecell}
\usepackage{verbatim}
\usepackage{acronym}
\usepackage[super]{nth}
\usepackage{tikz}
\usepackage{amsmath}
\usepackage{filecontents}
\usepackage[misc]{ifsym}
\usepackage{grffile}
\usepackage{tikz,pgf}
\usepackage{amsthm}
\usepackage[most]{tcolorbox}
\usepackage{enumitem}
\usepackage{trimclip}

\usetikzlibrary{calc}
\usetikzlibrary{positioning, fit, backgrounds, arrows.meta}
\makeatletter
\def\mfontsize{\f@size}

\newcommand{\F}{Fig.}

\newcommand{\T}{Table}

\usepackage{xcolor}

\definecolor{codegreen}{rgb}{0,0.6,0}
\definecolor{codegray}{rgb}{0.5,0.5,0.5}
\definecolor{codepurple}{rgb}{0.58,0,0.82}
\definecolor{backcolour}{rgb}{0.95,0.95,0.92}
\definecolor{cadmiumgreen}{rgb}{0.0, 0.42, 0.24}
\lstdefinestyle{mystyle}{
	escapeinside={(*@}{@*)},
	backgroundcolor=\color{backcolour},   
	commentstyle=\color{codegreen},
	keywordstyle=\color{magenta},
	numberstyle=\tiny\color{codegray},
	stringstyle=\color{codepurple},
	frame=shadowbox,
	basicstyle=\ttfamily\footnotesize,
	breakatwhitespace=false,         
	breaklines=true,                 
	captionpos=b,                    
	keepspaces=true,                 
	numbers=left,                    
	numbersep=5pt,                  
	showspaces=false,                
	showstringspaces=false,
	showtabs=false,                  
	tabsize=2,
}

\lstdefinestyle{gdb}
{
	backgroundcolor=\color{backcolour},
	keywordstyle=\color{magenta},
	escapeinside={(*@}{@*)},
	basicstyle=\scriptsize\color{black}\ttfamily,
	frame=shadowbox
}

\usepackage{subcaption}

\begin{document}

\title{GPU-Fuzz: Finding Memory Errors in Deep Learning Frameworks}

\author{Zihao Li}
\authornote{Zihao Li and Hongyi Lu contributed equally to this work.}
\affiliation{%
  \institution{Southern University of Science and Technology}
  \country{}
}
\email{lizh2025@mail.sustech.edu.cn}

\author{Hongyi Lu}
\authornotemark[1]
\affiliation{%
  \institution{Southern University of Science and Technology}
  \country{}
}
\email{luhy2017@mail.sustech.edu.cn}

\author{Yanan Guo}
\affiliation{%
  \institution{University of Rochester}
  \country{}
}
\email{yanan.guo@rochester.edu}

\author{Zhenkai Zhang}
\affiliation{%
  \institution{Clemson University}
  \country{}
}
\email{zhenkai@clemson.edu}

\author{Shuai Wang}
\affiliation{%
  \institution{Hong Kong University of Science and Technology}
  \country{}
}
\email{shuaiw@cse.ust.hk}

\author{Fengwei Zhang}
\authornote{Corresponding author.}
\affiliation{%
  \institution{Southern University of Science and Technology}
  \country{}
}
\email{zhangfw@sustech.edu.cn}

\begin{abstract}
  GPU memory errors are a critical threat to deep learning (DL) frameworks, leading to crashes or even security issues. We introduce \textsc{GPU-Fuzz}, a fuzzer locating these issues efficiently by modeling operator parameters as formal constraints. \textsc{GPU-Fuzz} utilizes a constraint solver to generate test cases that systematically probe error-prone boundary conditions in GPU kernels. Applied to PyTorch, TensorFlow, and PaddlePaddle, we uncovered 13 unknown bugs, demonstrating the effectiveness of \textsc{GPU-Fuzz} in finding memory errors.
  \end{abstract}
  
\maketitle

\section{Introduction}
\label{sec:intro}

GPUs are now an indispensable component of deep learning (DL) frameworks like PyTorch~\cite{paszke2019pytorch} and TensorFlow~\cite{abadi2016tensorflow}. However, the correctness of GPU computations is often threatened by memory corruptions, an insidious class of bugs stemming from the low-level CUDA kernels~\cite{nickolls2008scalable}. These errors, such as out-of-bounds access or misaligned memory addressing, can lead not only to system crashes but also to silent data corruption~\cite{fiala2012detection}, posing a significant threat to the reliability and security of AI applications.

However, locating these memory-related bugs remains a profound challenge. Existing fuzzers for DL systems are primarily designed to find arithmetic miscomputations in the DL compilers by generating diverse neural networks with different structures~\cite{liu2023nnsmith}. This approach, while effective for compiler testing, is ill-suited for uncovering memory errors due to the lack of exploration of the operator parameter space.

Our key insight is that uncovering GPU memory errors requires a shift in focus from the network structure to the operator parameters and memory layout. The precise combination of tensor shapes, data types, strides, and other parameters dictates the memory access patterns within a CUDA kernel. To effectively find memory bugs, a fuzzer must be able to reason about these intricate patterns and generate inputs that systematically explore the parameter space.

To this end, we designed and implemented \textsc{GPU-Fuzz}, a novel fuzzer specifically engineered to locate these memory-related bugs. Unlike existing DL fuzzers such as NNSmith~\cite{liu2023nnsmith}, \textsc{GPU-Fuzz} focuses on the operator level. It translates the complex semantic and memory-related rules of DL operators into formal constraints, which are then solved to generate diversified inputs that probe memory-related boundary conditions. This constraint-guided approach~\cite{godefroid2005dart} enables \textsc{GPU-Fuzz} to effectively stress-test the low-level CUDA kernels for memory safety.

The main contributions of this paper are as follows:
\begin{itemize}
    \item We propose a new fuzzing approach that targets GPU memory errors by systematically exploring the operator parameter space, a dimension orthogonal to existing DL fuzzers~\cite{liu2023nnsmith}.
    \item We design and implement \textsc{GPU-Fuzz}, a system that leverages constraint solving to automatically generate test cases that probe memory-related boundary conditions in low-level CUDA kernels.
    \item We demonstrate the effectiveness of \textsc{GPU-Fuzz} by uncovering 13 previously unknown bugs in major DL frameworks like PyTorch, TensorFlow, and PaddlePaddle.
\end{itemize}

\section{Background and Motivation}
\label{sec:bg}

\subsection{GPU Architecture}
Modern GPUs are massively parallel devices built on Streaming Multiprocessors (SMs), each executing hundreds of threads concurrently~\cite{nickolls2008scalable}.
This parallelism is supported by a complex memory hierarchy~\cite{kirk2016programming}.
However, harnessing this hardware for performance requires developers to manually manage data placement and movement~\cite{guide2020cuda}. 
The complexity of this manual task makes GPU kernels highly susceptible to memory errors such as out-of-bounds access and race conditions~\cite{kamath2021iguard}.

\begin{figure}[htbp]
	\centering
	\includegraphics[width=0.8\columnwidth]{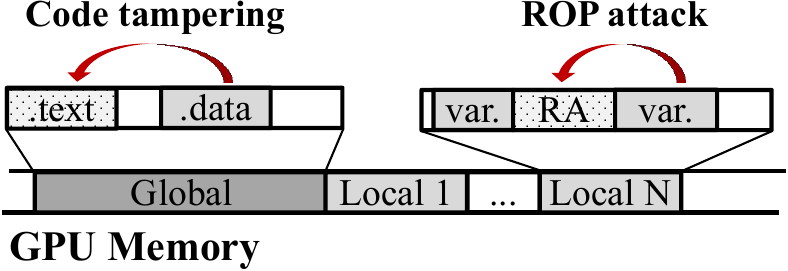}
	\caption{GPU memory layout and potential attacks.}
	\Description{Common attack vectors targeting GPU memory. (Left) Code tampering modifies kernel instructions in Global Memory. (Right) ROP attacks corrupt the return address (RA) on the thread-private stack in Local Memory.}
	\label{fig:gpu_attacks}
\end{figure}
As illustrated in \F~\ref{fig:gpu_attacks}, these memory errors can be exploited to compromise the kernel's integrity and control flow. The memory spaces most relevant to these attacks are:

\parh{Global Memory.}
The global memory is a large memory space shared by all threads. It is commonly used to store the kernel's executable instructions and global variables. Since previous works have shown that GPUs do not support W$\oplus$X permission~\cite{guo2024gpu}, an out-of-bounds write vulnerability can allow an attacker to modify the kernel's instructions residing in this shared space, fundamentally altering the program's intended logic.

\parh{Local Memory.}
Each thread possesses a private local memory space. This thread-local space stores function arguments, local variables, and control-flow data such as the return address. A stack-based buffer overflow, a prevalent type of memory error, can overwrite the saved return address. This vulnerability is the foundation for Return-Oriented Programming (ROP) attacks~\cite{guo2024gpu}, allowing an attacker to hijack the thread's control flow and divert execution to malicious payloads.

The existence of these vulnerabilities highlights the need for robust mechanisms to detect and mitigate memory corruption in GPU kernels.

\subsection{Deep Learning Operators}
Deep learning (DL) frameworks like PyTorch~\cite{paszke2019pytorch} and TensorFlow~\cite{abadi2016tensorflow} are built around a rich library of fundamental computational units called operators. 
These operators, such as convolution, pooling, matrix multiplication, and activation functions~\cite{chetlur2014cudnn}, serve as the elemental building blocks for constructing neural networks. 
While users interact with them through simple, high-level Python APIs, the underlying reality is far more complex. Each operator is backed by one or more highly-optimized, low-level programs known as kernels~\cite{chetlur2014cudnn}, which are executed on the GPU.

\begin{figure}[htbp]
	\centering
	\includegraphics[width=0.9\columnwidth]{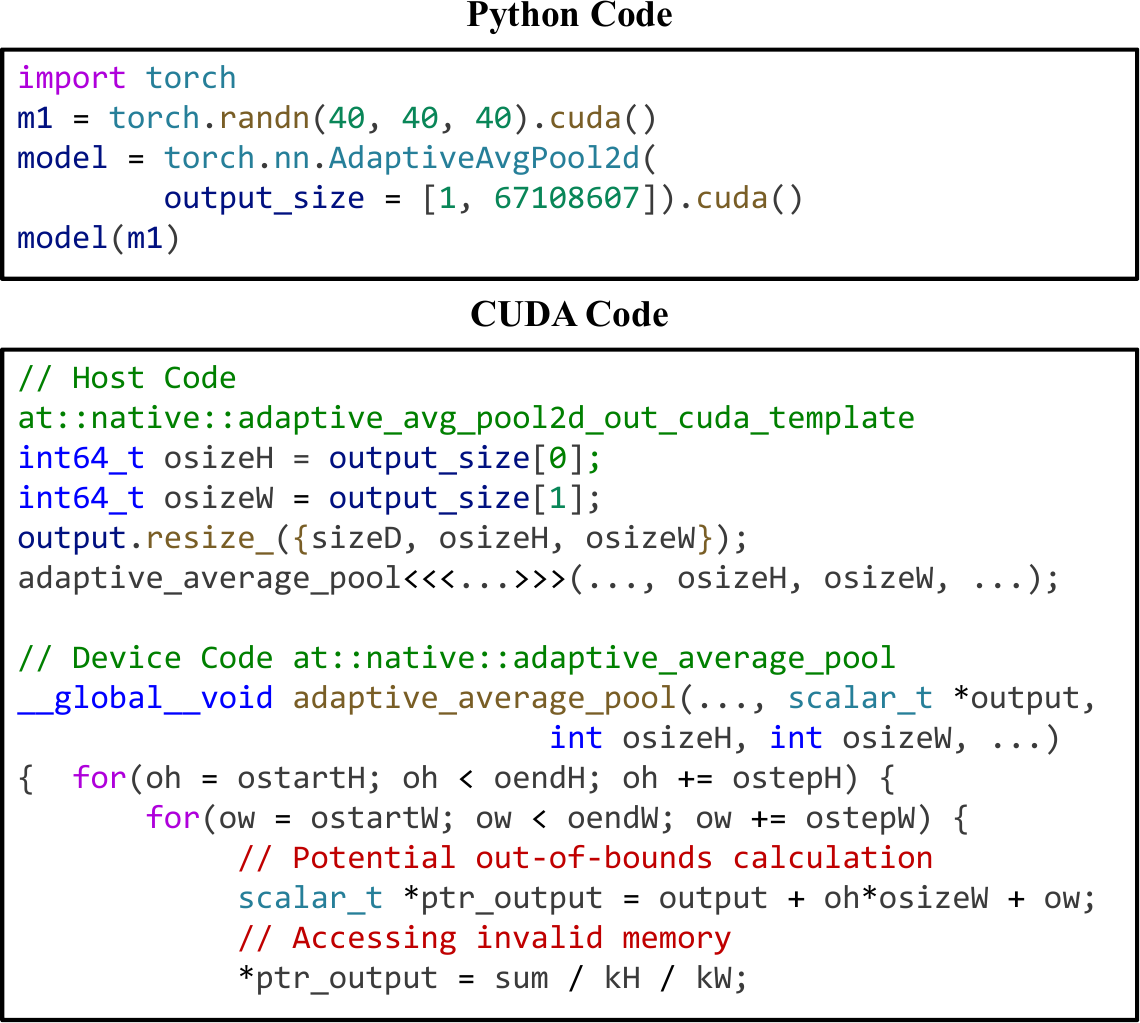}
	\caption{From Python API to CUDA kernel.}
	\Description{The figure illustrates how a PyTorch ConvTranspose2d operation with large dimensions triggers a CUDA kernel. Due to an integer overflow in grid dimension calculation, the CUDA kernel threads calculate indices that exceed the valid output buffer bounds, resulting in invalid global memory writes.}
	\label{fig:operatorcode}
\end{figure}

As illustrated in \F~\ref{fig:operatorcode}, high-level Python operator calls are translated into low-level CUDA kernel executions, where subtle implementation bugs can lead to various memory errors.

The complexity arises from the vast parameter space of each operator. 
A single convolution operator~\cite{dumoulin2016guide}, for instance, is governed by a multitude of parameters beyond the input tensor itself: kernel size, stride, padding, dilation, and channel groups. 
These parameters are not independent; they are bound by a complex set of semantic and mathematical constraints that dictate valid combinations and determine the output tensor's properties. 
To achieve state-of-the-art performance, framework developers implement these kernels manually in CUDA C++, employing sophisticated techniques like shared memory tiling and intricate pointer arithmetic to maximize data throughput~\cite{guide2020cuda}. 
This manual, performance-driven optimization often bypasses safer, high-level abstractions, making the kernel code a fertile ground for subtle memory errors~\cite{kamath2021iguard} that are triggered only by specific, often obscure, parameter configurations.

\subsection{Motivation}
GPU memory bugs represent a severe and often silent threat to the reliability and security of AI systems~\cite{papadimitriou2023silent}. 
These bugs can cause catastrophic failures in mission-critical applications like medical imaging~\cite{ronneberger2015u} and autonomous driving~\cite{lang2019pointpillars}, or be exploited for security attacks~\cite{pavlidakis2024guardian,park2021mind}. 

State-of-the-art DL fuzzers focus on the compiler stack, generating valid neural networks to find bugs~\cite{liu2023nnsmith,ma2023fuzzing}. 
This approach is ill-suited for finding low-level memory errors in GPU kernels. 
Such bugs are not typically triggered by network architecture, but by specific, often boundary-value, combinations of an operator's parameters (e.g., tensor shapes, strides). 
This leaves a fundamental blind spot: existing fuzzers like NNSmith~\cite{liu2023nnsmith}  do not systematically explore the intricate parameter space of individual operators where these memory bugs reside.

This observation reveals the need for a paradigm shift toward operator-level fuzzing. 
We introduce \textsc{GPU-Fuzz}, a system designed to explore the operator parameter space to uncover memory bugs in low-level CUDA kernels.

\section{System Design}
\label{sec:design}

\begin{figure}[htbp]
	\centering
	\vspace{-0.2em}
	\includegraphics[width=1\linewidth]{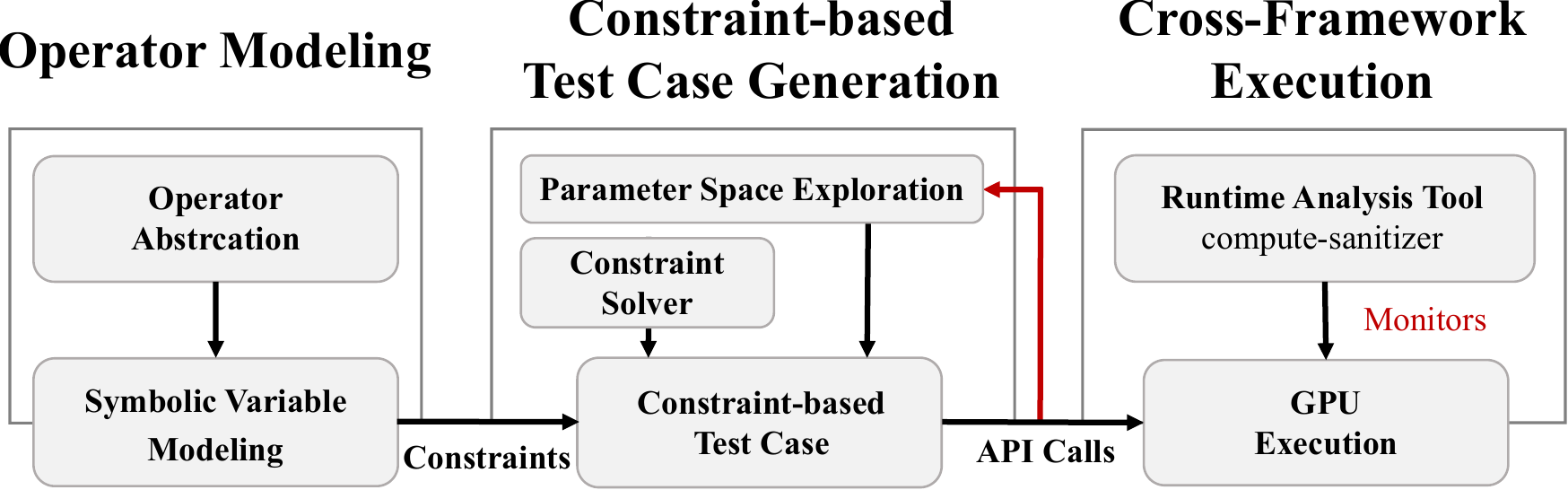}
	\vspace{-0.4em}
	\caption{The architecture of the \textsc{GPU-Fuzz} system.}
	\Description{A diagram illustrating the architecture of the GPU-Fuzz system. }
	\label{fig:arch}
	\vspace{-0.2em}
\end{figure}

This section details the architecture of \textsc{GPU-Fuzz}. As illustrated in \F~\ref{fig:arch}, the system consists of three main phases: operator modeling, constraint-based test case generation, and cross-framework execution.

\subsection{Operator modeling.}
\label{sec:modeling}

\textsc{GPU-Fuzz} models GPU operators through an abstraction layer that captures their parameter spaces and shape relationships. Each operator family (e.g., convolution, pooling) is represented by a unified model that defines the interface for input/output shapes and parameter constraints.

\begin{figure}[htbp]
	\centering
	\hspace*{-0.2cm}\includegraphics[width=0.9\columnwidth]{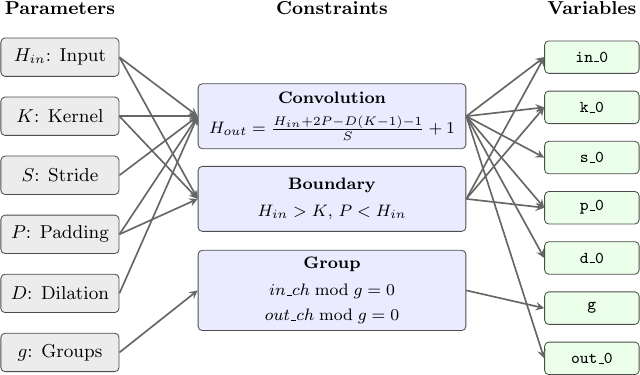}
	\caption{Constraint modeling for convolution operators.}
	\Description{A diagram illustrating the constraint modeling process for convolution operators.}
	\label{fig:constraint_modeling}
\end{figure}

As illustrated in \F~\ref{fig:constraint_modeling}, \textsc{GPU-Fuzz} encodes operator semantics into a set of constraint formulas with symbolic variables. 
Take the convolution operator in \F~\ref{fig:constraint_modeling} as an example, its core constraint formula~\cite{dumoulin2016guide} is $H_{out} = \frac{H_{in} + 2P - D(K-1) - 1}{S} + 1$, where $H_{in}$ and $H_{out}$ denote the input and output sizes, $P$ is padding, $D$ is dilation, $K$ is kernel size, and $S$ is stride. 
Only after an input satisfies the constraint formulas, can the operator be correctly instantiated and executed. 
Aside from this core constraint, there are also various additional constraints (e.g., $H_{in} > K$) to ensure the semantic correctness of the operator. 
This constraint modeling approach allows us to effectively generate valid test cases for DL operators.

To ensure correctness, we manually extract the constraint formulas from the documentation of each operator. It takes approximately a month for two authors to extract the constraint formulas from the documentation of each operator independently and then cross-check their results. 
These two authors are experts in the field of deep learning and have a deep understanding of the semantics of each operator, which ensures the correctness of \textsc{GPU-Fuzz} to a great extent. In total, we extracted 45 constraints for 13 operators.

\subsection{Constraint-based Test Case Generation}
\label{sec:testcase}

\begin{figure}[htbp]
	\centering
	\includegraphics[width=\columnwidth]{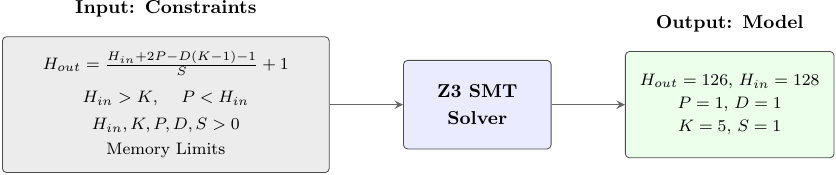}
	\caption{Constraint solving process.}
	\Description{A diagram illustrating how constraints are solved using Z3 SMT solver to generate concrete parameter values.}
	\label{fig:constraint_solving}
\end{figure}

\parh{Constraint solving.}
Once an operator is modeled according to its constraints, \textsc{GPU-Fuzz} employs Z3's SMT solver~\cite{de2008z3} to find a satisfying assignment for all symbolic variables. 
As illustrated in \F~\ref{fig:constraint_solving}, to generate a test case for a convolution operator, the solver must find concrete values for $H_{out}$, $H_{in}$, $P$, $D$, $K$, and $S$ that satisfy the set of constraints shown in the figure. 
In the case of \F~\ref{fig:constraint_solving}, a possible solution is $H_{out}=126$, $H_{in}=128$, $P=1$, $D=1$, $K=5$, and $S=1$.

Although SMT solvers like Z3~\cite{de2008z3} effectively find a solution for the given constraints, they are not designed to systematically explore the entire parameter space. Specifically, solvers like Z3~\cite{de2008z3} tend to return a single boundary solution for the symbolic variables~(e.g., $x=0$ for $x\geq 0$), which is insufficient for a comprehensive fuzzing campaign. 
To address this limitation, previous works like NNSmith~\cite{liu2023nnsmith} force Z3~\cite{de2008z3} to return multiple solutions by adding a periodic perturbation to the original constraint~(e.g., $x>2^n$).
\textsc{GPU-Fuzz}, on the other hand, employs a novel constraint-guided search strategy to systematically explore the parameter space.

\begin{figure}[htbp]
	\centering
	\includegraphics[width=1\linewidth]{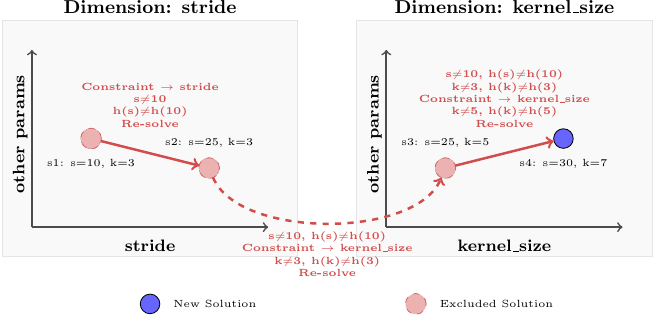}
	\caption{An example of the iterative parameter space exploration. At each step, a parameter is randomly selected and a new constraint that excludes its current value is incrementally added to guide the search for the next solution.}
	\Description{A diagram illustrating the iterative parameter space exploration process.}
	\label{fig:exploration}
\end{figure}

\parh{Parameter space exploration.}
To systematically explore the parameter space, \textsc{GPU-Fuzz} employs an iterative constraint-guided search strategy, as illustrated in \F~\ref{fig:exploration}. The process begins with an initial solution, $s1$ (e.g., $stride=10, kernel\_size=3$), found by the solver. At each iteration, the system randomly selects a parameter dimension to constrain. For instance, it might first select $stride$, adding constraints to exclude its current value, which guides the solver to a new solution like $s2$. In the next step, the system could randomly select $kernel\_size$, accumulating new constraints (e.g., $kernel\_size\neq 3$ and $h(kernel\_size)\neq h(3)$) upon the existing ones. This compels the solver to navigate towards unexplored regions, yielding another solution like $s3$. 

Notably, to enhance both solver efficiency and solution diversity, we propose to incorporate not only direct value exclusion (e.g., $stride\neq 10$) but also hash-based constraints (e.g., $h(stride)\neq h(10)$). 
The hash function transforms input values through a series of bit-mixing operations that introduce dispersion properties. 
Specifically, the function applies alternating right-shift, XOR, and multiplication operations to ensure that even small changes in input values result in different hash values, effectively avoiding identity mapping and poor distribution. 
The hash-based constraint complements the direct value exclusion by preventing the solver from exploring similar regions, which significantly improves solution diversity.

This incremental strategy ensures that \textsc{GPU-Fuzz} can continuously and efficiently generate diverse test cases.

\subsection{Cross-framework Execution}
\label{sec:execution}

\begin{figure}[htbp]
	\centering
	\vspace{-0.2em}
	\includegraphics[width=\columnwidth]{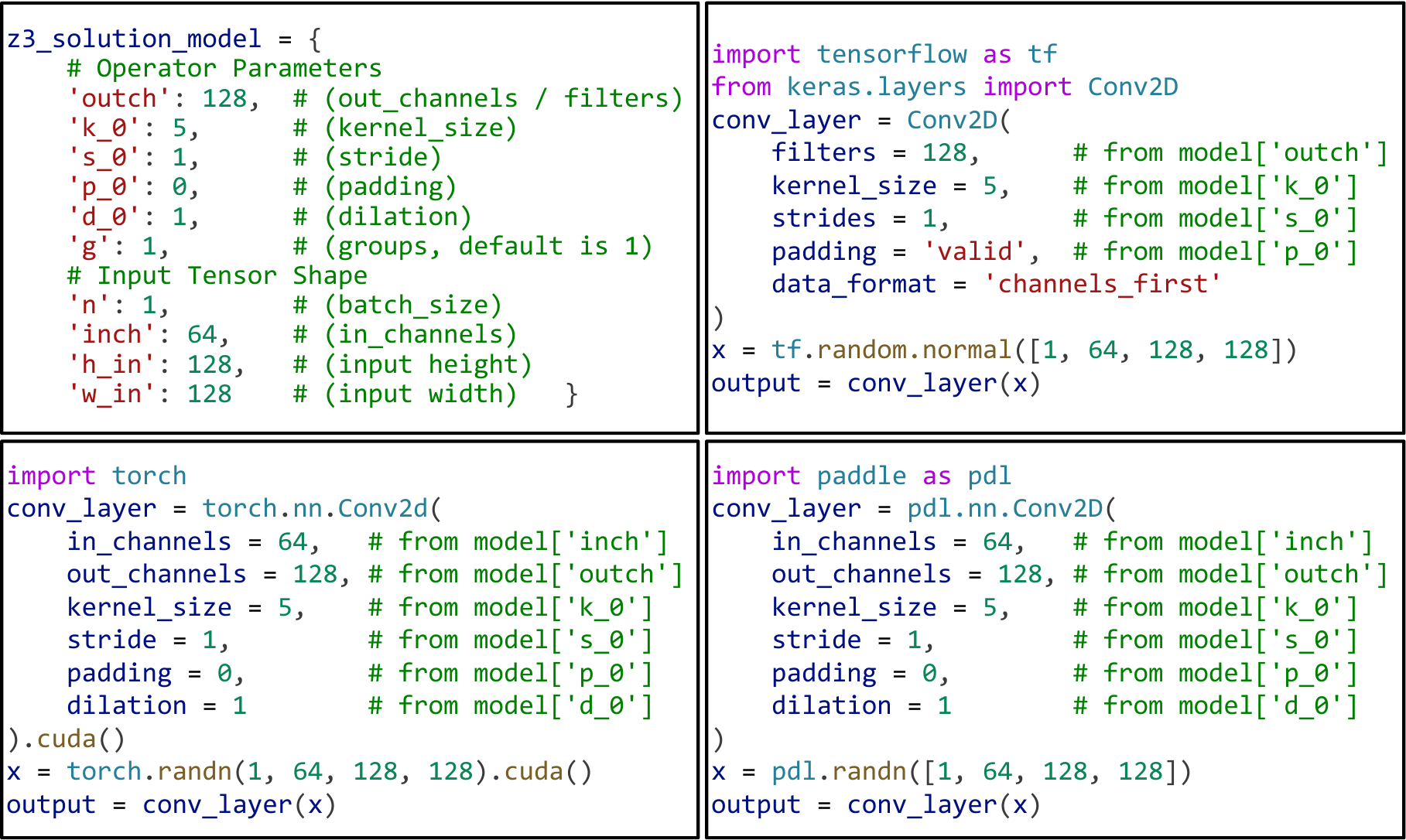}
	\vspace{-0.4em}
	\caption{Cross-framework materialization example.}
	\Description{A diagram illustrating how a generic model's parameters are materialized into framework-specific API calls for PyTorch, PaddlePaddle, and TensorFlow.}
	\label{fig:execution}
	\vspace{-0.2em}
\end{figure}

Once a valid set of parameters is generated, \textsc{GPU-Fuzz} executes it as a test case across multiple deep learning frameworks, including PyTorch~\cite{paszke2019pytorch}, PaddlePaddle~\cite{ma2019paddlepaddle}, and TensorFlow~\cite{abadi2016tensorflow}, to detect GPU memory bugs. 
As illustrated in \F~\ref{fig:execution}, this process translates the abstract, framework-agnostic parameters from the solver into concrete parameters with specific API calls. 
For instance, the generic \texttt{outch} parameter is mapped to \texttt{out\_channels} in PyTorch and \texttt{filters} in TensorFlow. 
To detect memory errors and kernel failures, each execution is wrapped by NVIDIA's compute-sanitizer~\cite{nvidia2023compsan}. 
This approach significantly improves the detection of memory errors.

\section{Implementation}
\label{sec:impl}
This section details the implementation of the system design described in Section~\ref{sec:design}. Our system is developed in Python and comprises 2,628 lines of code.
    
We implemented a library where each operator is represented as a distinct class to realize the operator modeling and constraint-based test case generation concepts described in Section~\ref{sec:design}. This modeling approach was applied to 13 operators (\T~\ref{tab:operators}), chosen for their prevalence in deep learning models and their complex, error-prone memory access patterns.

The translation process (Section~\ref{sec:execution}) is implemented to translate the generated parameter values into framework-specific API calls. When compute-sanitizer~\cite{nvidia2023compsan} detects an error during execution, the system automatically archives the execution logs to ensure reproducibility.

\begin{table}[t]
    \centering
    \caption{Supported Operators in \textsc{GPU-Fuzz}}
    \label{tab:operators}
    \footnotesize
    \begin{tabular}{@{}lp{5.0cm}@{}}
    \toprule
    \textbf{Operator Family} & \textbf{Specific Operators} \\
    \midrule
    Convolution & \makecell[l]{Conv (1d, 2d, 3d), ConvTranspose (1d, 2d, 3d)} \\
    \midrule
    Pooling & \makecell[l]{MaxPool (1d, 2d, 3d), AvgPool (1d, 2d, 3d), \\ FractionalMaxPool (2d, 3d), LPPool (1d, 2d, 3d), \\ AdaptiveAvgPool (1d, 2d, 3d),\\ AdaptiveMaxPool (1d, 2d, 3d)} \\
    \midrule
    Padding & \makecell[l]{ReflectionPad (1d, 2d, 3d),\\ ReplicationPad (1d, 2d, 3d),\\ ZeroPad (1d, 2d, 3d), ConstantPad (1d, 2d, 3d),\\ CircularPad (1d, 2d, 3d)} \\
    \midrule
    Element-Wise Unary & \makecell[l]{Activation: ELU, ReLU, GELU, Sigmoid, Tanh, ...\\ Arithmetic: abs, sin, cos, sqrt, exp, log, ...} \\
    \midrule
    Element-Wise Binary & \makecell[l]{add, sub, mul, div, pow, remainder,\\ logaddexp, atan2, ...} \\
    \midrule
    Matrix Ops & MatMul, BMM \\
    \midrule
    Concat & cat, concat \\
    \bottomrule
    \end{tabular}
    \end{table}
    \vspace{-0.4em}
\section{Evaluation}
\label{sec:eval}

In this section, we evaluate the effectiveness and efficiency of \textsc{GPU-Fuzz}. We aim to answer the following research questions (RQs):
\begin{itemize}[leftmargin=1em]
    \item \textbf{RQ1:} How effective is \textsc{GPU-Fuzz} in uncovering real-world bugs in major deep learning frameworks?
    \item \textbf{RQ2:} How does \textsc{GPU-Fuzz} compare with state-of-the-art DL fuzzers in terms of test case generation and bug discovery, particularly for GPU memory errors?
\end{itemize}

\subsection{Experimental Setup}
\label{sec:setup}
All experiments were conducted on a server with the configuration detailed in \T~\ref{tbl:setup}. We established isolated Conda environments for each of the three target frameworks (PyTorch, TensorFlow, and PaddlePaddle) to manage their specific dependencies. The core hardware, operating system, and NVIDIA driver were consistent across all tests.

\begin{table}[h!]
\centering
\caption{Experimental Environment Configuration.}
\label{tbl:setup}
\footnotesize
\begin{tabular}{@{}ll@{}}
\toprule
\textbf{Component} & \textbf{Specification} \\ \midrule
\multicolumn{2}{@{}l}{\textbf{Hardware}} \\
\hspace{1em}CPU & 2 x Intel Xeon Silver 4510 \\
\hspace{1em}GPU & NVIDIA H100 PCIe \\
\midrule
\multicolumn{2}{@{}l}{\textbf{Software}} \\
\hspace{1em}Operating System & Ubuntu 24.04.2 LTS \\
\hspace{1em}NVIDIA Driver & 580.82.07 \\
\hspace{1em}CUDA Runtime & 12.8.93 \\
\hspace{1em}Python & 3.11.13 \\
\midrule
\multicolumn{2}{@{}l}{\textbf{Frameworks}} \\
\hspace{1em}PyTorch & PyTorch 2.3.1+cu121 with cuDNN 8.9.2.26 \\
\hspace{1em}TensorFlow & 2.20.0, with cuDNN 9.13.0.50 \\
\hspace{1em}PaddlePaddle & 2.6.1 \\
\bottomrule
\end{tabular}%
\end{table}

\subsection{Bug Discovery Effectiveness}

Over the course of our evaluation, \textsc{GPU-Fuzz} uncovered a total of 13 previously unknown bugs across the three frameworks. \T~\ref{tbl:bugs} presents a summary of these findings. The bugs span a range of failure modes, from low-level memory corruption to API-level exceptions. 
We identified 7 distinct memory access violations (e.g., out-of-bounds or misaligned writes). Among these, 5 were silent memory corruptions that do not trigger any API-level crash and are only detectable with specialized tools like compute-sanitizer~\cite{nvidia2023compsan}. 

\begin{table*}[htbp]
\centering
\caption{Summary of Bugs Discovered by \textsc{GPU-Fuzz}.}
\label{tbl:bugs}
\resizebox{\textwidth}{!}{%
\begin{tabular}{@{}lllllll@{}}
\toprule
\textbf{ID} & \textbf{Framework} & \textbf{Operator} & \textbf{Bug Type} & \textbf{Root Cause} & \textbf{Failure Mode} & \textbf{Status} \\ \midrule
\textbf{Bug\textsubscript{1}} & PyTorch & \texttt{conv\_transpose2d} & OOB Global Write & Incorrect grid dimension calculation & GPU-Level Exception (CUBLAS) & Confirmed \\
\textbf{Bug\textsubscript{2}} & PyTorch & \texttt{bmm\_sparse} & Misaligned Global Write & Incorrect pointer arithmetic in CUSPARSE & Silent Memory Corruption & Reported \\
\textbf{Bug\textsubscript{3}} & PyTorch & \texttt{adaptive\_avg\_pool2d} & OOB Global Write & Flawed boundary checks in CUDA kernel & Silent Memory Corruption & Confirmed \\
\textbf{Bug\textsubscript{4}} & PyTorch & \texttt{replication\_pad2d} & OOB Global Write & Incorrect grid dimension calculation & Silent Memory Corruption & Confirmed \\
\textbf{Bug\textsubscript{5}} & PyTorch & \texttt{adaptive\_max\_pool2d} & OOB Global Write & Flawed boundary checks in CUDA kernel & Silent Memory Corruption & Confirmed \\
\textbf{Bug\textsubscript{6}} & PyTorch & \texttt{conv\_transpose3d} & OOB Shared Read & Incorrect index calculation for shared memory & Silent Memory Corruption & Fixed \\
\textbf{Bug\textsubscript{7}} & PyTorch & \texttt{reflection\_pad1d} & Invalid Launch Config & Integer overflow in \texttt{torch.compile} symint logic & GPU-Level Exception (CUDA) & Confirmed \\
\textbf{Bug\textsubscript{8}} & TensorFlow & \texttt{Conv2D} & OOB Global Read & Incorrect index calculation in kernel & Silent Read / Downstream Crash & Confirmed \\
\textbf{Bug\textsubscript{9}} & TensorFlow & \texttt{Conv2D} & Integer Overflow & Overflow in launch config calculation & CPU-Side Assert & Confirmed \\
\textbf{Bug\textsubscript{10}} & PaddlePaddle & \texttt{conv2d\_transpose} & Precondition Violation & Integer overflow in tensor dimension calculation & CPU-Side Assert & Confirmed \\
\textbf{Bug\textsubscript{11}} & PaddlePaddle & \texttt{conv3d\_transpose} & Illegal Instruction & Invalid parameters passed to cuDNN kernel & GPU-Level Exception (cuDNN) & Confirmed \\
\textbf{Bug\textsubscript{12}} & PaddlePaddle & \texttt{conv2d\_transpose} & Bad API Parameter & Invalid parameter combination passed to cuDNN & GPU-Level Exception (cuDNN) & Confirmed \\
\textbf{Bug\textsubscript{13}} & PaddlePaddle & \texttt{conv2d\_transpose} & Invalid Launch Config & Incorrect grid/block dimension calculation & GPU-Level Exception (CUDA) & Confirmed \\
\bottomrule
\end{tabular}
}
\end{table*}

\parh{Bug Patterns.} 
Our findings reveal important patterns across three distinct failure modes:
\begin{itemize}[leftmargin=1em]
    \item \textbf{Silent Memory Corruption:} The most critical category, where out-of-bounds or misaligned memory access occurs without causing any API-level error. These are the most insidious bugs as they can lead to silent data corruption and are only detectable with low-level memory debuggers.
    \item \textbf{GPU-Level Exceptions:} The second category, where invalid parameters or configurations cause CUDA, cuDNN, or CUBLAS libraries to return an error, which is then typically caught and reported by the framework.
    \item \textbf{CPU-Side Asserts:} The final category, where issues like integer overflows occur on the CPU during the calculation of kernel parameters, preventing the GPU launch altogether.
\end{itemize}
\noindent A common root cause across all frameworks was incorrect grid dimension calculations or flawed boundary checks, with transposed convolutions being particularly error-prone. 

\subsection{Comparative Study}
\label{sec:comparison}
To quantitatively validate our approach, we conducted a comparative study against NNSmith~\cite{liu2023nnsmith}, a state-of-the-art DL fuzzer. We conducted five independent 4-hour fuzzing runs for each tool on the same hardware targeting PyTorch, with both tools running in identical Conda environments.

\T~\ref{tbl:comparison} summarizes the results. NNSmith generated on average $19{,}063\pm 360$ test cases and uncovered $296\pm 19$ bugs. Most of its findings are numerical mismatches rather than memory-safety issues. In contrast, \textsc{GPU-Fuzz} generated on average $51{,}860\pm 1{,}559$ test cases and uncovered $106\pm 8$ real bugs excluding out-of-memory errors, including $26\pm 5$ critical memory errors and $80\pm 7$ configuration errors. These memory errors represent severe security vulnerabilities that could result in data corruption, information leakage, or system crashes.

\begin{table}[htbp]
\centering
\caption{Comparative Results}
\label{tbl:comparison}
\footnotesize
\begin{tabular}{@{}lcc@{}}
\toprule
\textbf{Metric} & \textbf{NNSmith} & \textbf{GPU-Fuzz} \\
\midrule
\textbf{Test Cases Generated} & $19{,}063\ \pm\ 360$ & $51{,}860\ \pm\ 1{,}559$ \\
\textbf{Total Bugs}\textsuperscript{*} & $296\ \pm\ 19$ & $106\ \pm\ 8$ \\
\midrule
\multicolumn{3}{@{}l}{\textbf{Bug Breakdown by Type}} \\
\hspace{1em}Memory Errors & 0 & $26\ \pm\ 5$ \\
\hspace{1em}Configuration Errors & 0 & $80\ \pm\ 7$ \\
\hspace{1em}Inconsistencies & $293\ \pm\ 19$ & 0 \\
\hspace{1em}Exceptions & $3\ \pm\ 1$ & 0 \\
\midrule
\textbf{Runtime} & \multicolumn{2}{c}{4 GPU-hours each} \\
\bottomrule
\multicolumn{3}{@{}l}{\scriptsize\textsuperscript{*}GPU-Fuzz total excludes out-of-memory errors.} \\
\multicolumn{3}{@{}l}{\scriptsize\textsuperscript{**}NNSmith and GPU-Fuzz results are mean $\pm$ std over 5 independent runs.} \\
\end{tabular}%
\end{table}

\parh{Key Findings.} 
Our analysis reveals two critical insights. First, \textsc{GPU-Fuzz} generated nearly three times more test cases than NNSmith, demonstrating the efficiency of constraint-guided parameter fuzzing in systematically exploring the operator parameter space. Second, \textsc{GPU-Fuzz} uncovered $26\pm 5$ memory errors that pose security risks, while NNSmith's findings were primarily numerical precision issues that typically do not threaten memory safety. This demonstrates that \textsc{GPU-Fuzz} addresses a blind spot in GPU memory security testing that existing DL fuzzers frequently miss. The two approaches are complementary: NNSmith excels at uncovering compiler-related bugs and numerical inconsistencies, while \textsc{GPU-Fuzz} fills the gap in GPU memory security testing at the operator parameter level.

\subsection{Case Study}
\label{sec:case_study}
\begin{figure}[htbp]
\centering
\vspace{-0.2em}
\includegraphics[width=0.9\columnwidth]{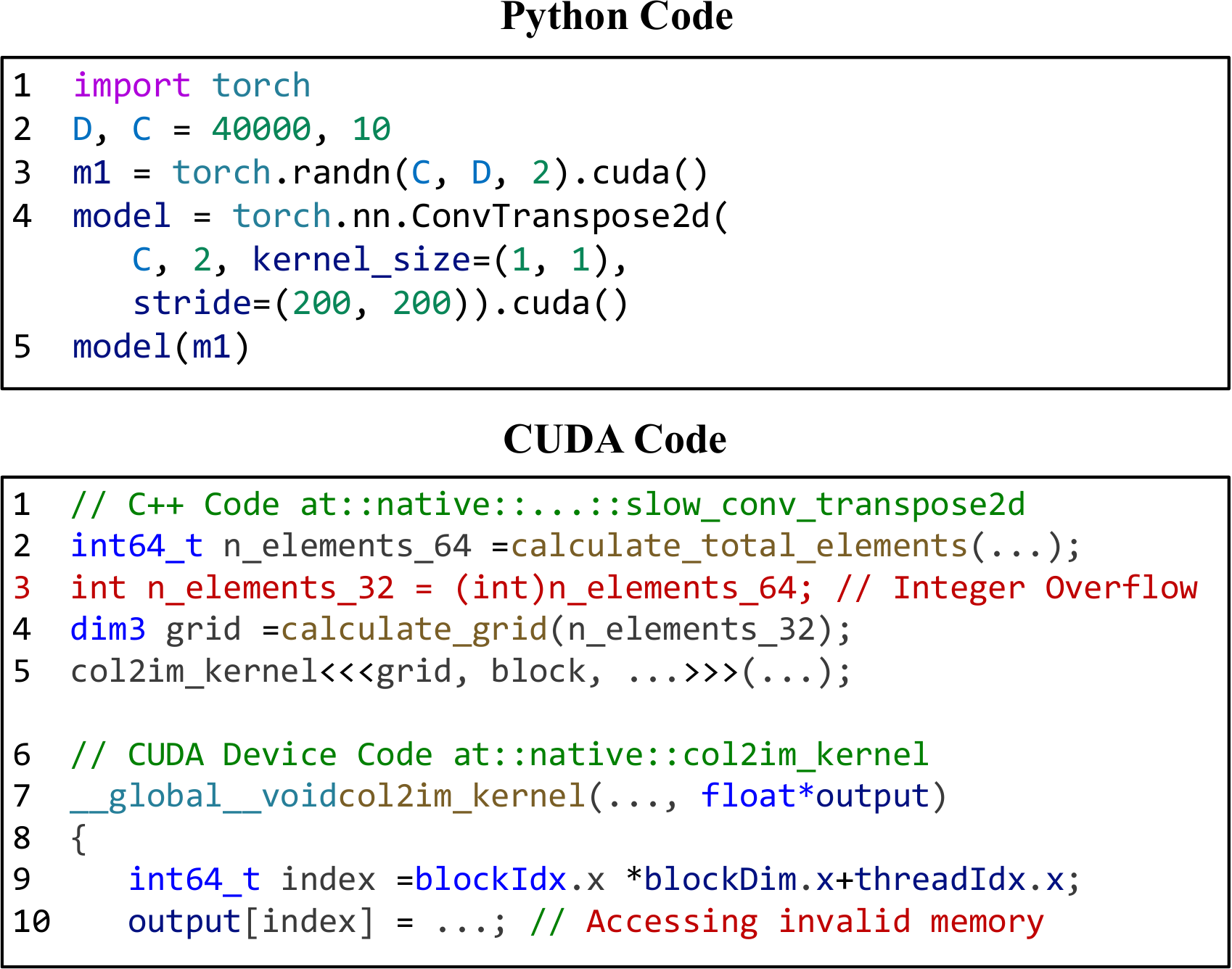}
\vspace{-0.4em}
\caption{A minimal PoC in Python and the corresponding CUDA implementation that triggers a memory bug in PyTorch's \texttt{ConvTranspose2d}.}
\Description{A code snippet showing a minimal proof-of-concept in Python that triggers a memory corruption bug in PyTorch's ConvTranspose2d operator, along with the corresponding CUDA code demonstrating an integer overflow when calculating the total number of elements, leading to invalid memory access.}
\label{fig:poc_code}
\vspace{-0.2em}
\end{figure}

To illustrate the practical utility of \textsc{GPU-Fuzz}, we present a minimal proof-of-concept (PoC) for a memory corruption bug uncovered in PyTorch's \texttt{ConvTranspose2d} operator. The PoC is shown in \F~\ref{fig:poc_code}.

The key to triggering this bug lies in the parameter combination automatically generated by our fuzzer, particularly the extremely large stride value of $(200, 200)$ combined with input dimensions of $(10, 40000, 2)$. 
While these values are semantically valid according to PyTorch's API, they represent a corner case that is unlikely to be covered by manual tests. 
As illustrated in \F~\ref{fig:poc_code}, the root cause is an integer overflow in the C++ host code: when calculating the total number of elements for the CUDA kernel, a 64-bit integer value is cast to a 32-bit integer, which causes truncation. 
This overflowed value is then used to calculate the CUDA grid dimensions, resulting in an undersized grid that cannot cover all required memory operations. 
When the \texttt{col2im\_kernel} executes, threads calculate 64-bit indices that exceed the actual allocated buffer size, leading to out-of-bounds memory writes. 

\begin{figure}[htbp]
    \centering
    \vspace{-0.2em}
    \includegraphics[width=0.8\columnwidth]{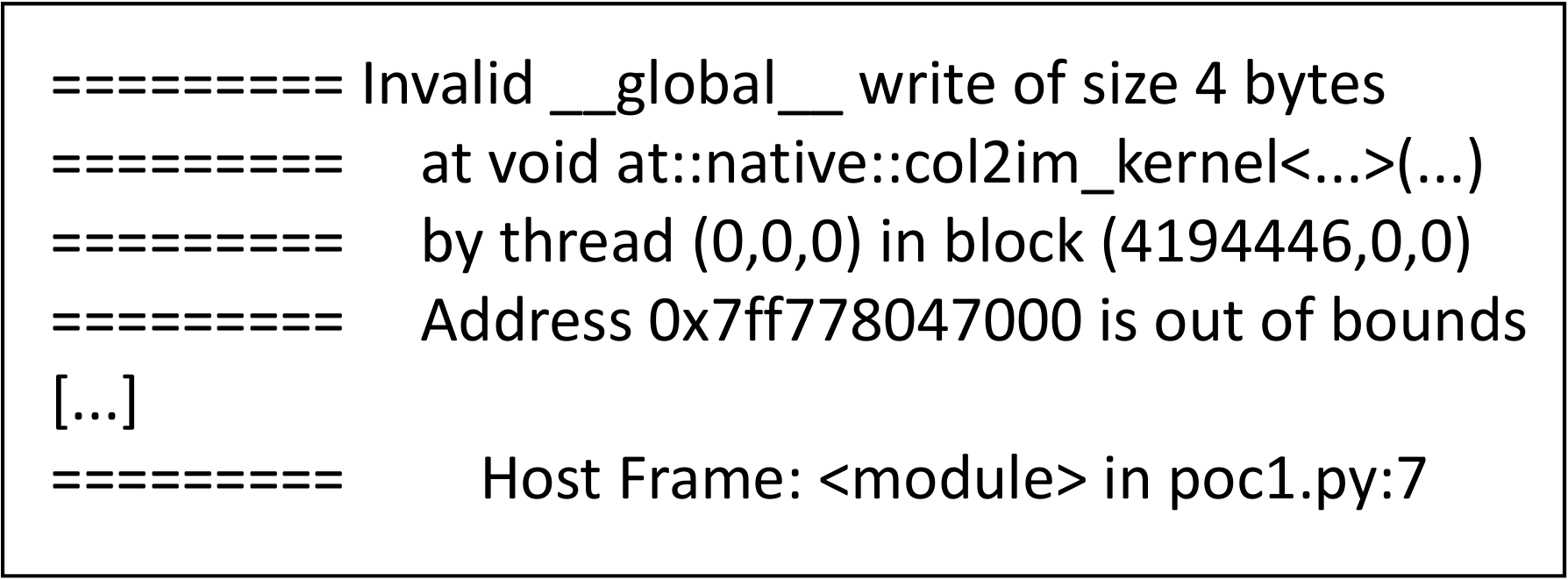}
    \vspace{-0.4em}
    \caption{The error message for the PoC.}
    \Description{An error log showing the error message for the proof-of-concept code in the previous figure.}
    \label{fig:poc_log}
    \vspace{-0.2em}
    \end{figure}
    
As shown in \F~\ref{fig:poc_log}, compute-sanitizer detects that a CUDA thread attempts to write to an out-of-bounds address, which is before the nearest valid allocation. 
This type of silent memory corruption is a severe bug that can lead to incorrect results or unpredictable behavior without causing an explicit Python crash, and demonstrates the ability of \textsc{GPU-Fuzz} to systematically uncover severe, hidden bugs in mature deep learning frameworks by exploring non-trivial parameter spaces.

\section{Discussion}
\label{sec:discussion}
While our results demonstrate the effectiveness of \textsc{GPU-Fuzz}, we acknowledge several limitations and areas for future work.

\parh{Manual Modeling Effort.} 
The quality of \textsc{GPU-Fuzz} depends on its constraint library. Our current library supports 13 operators, but this is a subset of the hundreds of operators available. Extending coverage requires manual effort to model constraints (100-150 LoC per family). Future work could explore semi-automating constraint extraction from framework documentation to improve scalability.

\parh{Limited Oracle.} 
Our primary oracle, compute-sanitizer, excels at finding memory errors but cannot detect silent numerical correctness issues or performance regressions. Differential fuzzing against a trusted CPU implementation is a promising direction for a more comprehensive oracle.

\section{Related Work}
\label{sec:related_work}
Fuzzing for DL systems has received continuous attention over the past years~\cite{shen2021comprehensive}.
To capture bugs from complicated modern DL compiler stacks, one of the effective approaches is generating valid neural network models. Works like NNSmith~\cite{liu2023nnsmith}, TVMFuzz~\cite{shen2021comprehensive}, and HirGen~\cite{ma2023fuzzing} pioneered this direction, proving effective at identifying compiler-related bugs, such as graph-level optimization issues and IR transformation errors.
Following this, other works propose testing different components of the DL stack. LEMON~\cite{wang2020deep}, for example, tests DL library implementations by generating model variants to detect inconsistencies across different libraries. More recent studies, like Orion~\cite{harzevili2025history}, specifically focus on the API layer by generating test inputs guided by historical bug patterns.
While effective, these approaches primarily operate at higher abstraction levels. They are not designed to systematically probe the low-level memory access patterns within the GPU kernels that execute the operators. Testing GPU kernels, in general, has its own line of research. GPUVerify~\cite{betts2012gpuverify} employs formal methods to verify kernel correctness by detecting data races and barrier divergence. DeepSmith~\cite{cummins2018compiler} generates random CUDA programs from scratch to stress-test the CUDA compiler~\cite{guide2020cuda} itself.

In contrast, \textsc{GPU-Fuzz} differs from this previous work. Instead of generating entire models like NNSmith~\cite{liu2023nnsmith} or entire CUDA programs like DeepSmith~\cite{cummins2018compiler}, it focuses on parameter space fuzzing of existing DL operators to uncover bugs at the kernel level.

\section{Responsible Disclosure}
We have disclosed all 13 discovered bugs responsibly to the respective development teams of PyTorch~\cite{paszke2019pytorch}, TensorFlow~\cite{abadi2016tensorflow}, and PaddlePaddle~\cite{ma2019paddlepaddle} by opening detailed issue reports in their official code repositories. 
At the time of writing, several of these issues have been acknowledged by the developers. 
We are committed to collaborating with the framework maintainers to help improve the security and robustness of the deep learning ecosystem.

\section{Conclusion}
\label{sec:conclusion}
We introduced \textsc{GPU-Fuzz}, a constraint-guided fuzzer that aims to find memory errors in deep learning operators. By shifting the focus from network models to the operator parameter space, \textsc{GPU-Fuzz} explores boundary conditions that may trigger low-level vulnerabilities. 
Our approach was able to uncover 13 previously unknown bugs in widely used frameworks such as PyTorch, TensorFlow, and PaddlePaddle, most of which were silent memory errors.  
This work suggests that securing modern AI systems may benefit from a complementary strategy combining both model and operator parameter space fuzzing. We hope that by making our tool and findings publicly available, we can contribute to improving the reliability and security of these foundational technologies.

\begin{acks}
We would like to thank the COMPASS members for their insightful comments. This work is partly supported by the National Natural Science Foundation of China under Grant No. U2541211, No. 62372218, and No. U24A6009. This work was also supported in part by a grant from the Research Grants Council of the Hong Kong Special Administrative Region, China HKUST C6004-25G and an ITF grant under the contract ITS/161/24FP. Dr. Yanan Guo and Dr. Zhenkai Zhang are supported by separate research gifts from HydroX AI, respectively.
\end{acks}

\bibliographystyle{ACM-Reference-Format}
\bibliography{ref}

\appendix

\end{document}